\documentclass[aps,prl,singlecolumn]{revtex4}
\usepackage{graphicx}
\usepackage{amsmath,bbold}
\usepackage{amssymb}
\usepackage{braket}
\usepackage{hyperref}
\usepackage{subfigure}
\usepackage[usenames,dvipsnames]{color}

\usepackage{graphicx}
\usepackage{subfigure}
\usepackage{slashed}
\usepackage{relsize}

\newcommand{\ct}{\cite}
\newcommand{\bi}{\bibitem}
\newcommand{\beq}{\begin{equation}}
\newcommand{\eeq}{\end{equation}}
\newcommand{\bea}{\begin{eqnarray}}
\newcommand{\eea}{\end{eqnarray}}
\newcommand{\al}{\alpha}
\newcommand{\de}{\delta}
\newcommand{\ep}{\epsilon}
\newcommand{\ga}{\gamma}
\newcommand{\Ga}{\Gamma}

\newcommand{\si}{\sigma}
\newcommand{\om}{\omega}
\newcommand{\da}{\dagger}
\newcommand{\non}{\nonumber}

\newcommand{\fl}[1]{\lfloor #1 \rfloor}
\newcommand{\flb}[1]{\Bigg\lfloor #1 \Bigg\rfloor}
\newcommand{\ta}{\theta}
\newcommand{\fll}[1]{\big\lfloor #1 \big\rfloor}

\begin{document}

\title{Fibonacci steady states in a driven integrable quantum system}
\author{Somnath Maity$^1$, Utso Bhattacharya$^1$, Amit Dutta$^1$ and
Diptiman Sen$^2$}
\affiliation{$^1$Department of Physics, Indian Institute of Technology,
Kanpur 208016, India \\
$^2$Centre for High Energy Physics, Indian Institute of Science, Bengaluru
560012, India}

\begin{abstract}
We study an integrable system that is reducible to free fermions by a 
Jordan-Wigner transformation which is subjected to a Fibonacci driving protocol
based on two non-commuting Hamiltonians. In the high frequency limit $\om \to 
\infty$, we show that the system reaches a nonequilibrium steady state, up to
some small fluctuations which can be quantified. For each momentum $k$, the 
trajectory of the stroboscopically observed state lies between two concentric 
circles on the Bloch sphere; the circles represent the boundaries of the small 
fluctuations. The residual energy is found to oscillate in a quasiperiodic way 
between two values which correspond to the two Hamiltonians that define the 
Fibonacci protocol. These results can be understood in terms of an effective 
Hamiltonian which simulates the dynamics of the system in the high frequency 
limit.
\end{abstract}

\maketitle

Through recent experimental progress~\cite{bloch08,lewenstein12,jotzu14,
greiner02,kinoshita06}, it has been realized that whether the unitary time 
evolution of a closed many-body quantum system in the thermodynamic limit
leads to a Gibbs ensemble after an asymptotically long time depends on the 
nature of the system and the initial state under consideration. 
To address this question, one considers a small subsystem of the entire 
system while the rest of the system acts as a bath. A system is said to 
thermalize when the long-time equilibrium properties of the subsystem are
correctly represented by considering a canonical (or grand canonical) ensemble 
for the whole system. In such a scenario, the system respects the eigenstate 
thermalization hypothesis~\cite{deutsch91,srednicki94,rigol08}. The usual 
quantum statistical mechanics then holds and can be applied successfully to 
understand the long-time steady states of the subsystem. However, many-body 
localized systems~\cite{pal10,nandkishore15} are examples where a quantum 
many-body system does not thermalize under unitary dynamics and retains the 
memory of the initial state. {\it Integrable} closed many-body quantum systems 
provide another example where the eigenstate thermalization hypothesis is 
violated although the entropy maximization principle still remains valid and 
an appropriate consideration of the extensive number of conservation laws
usually leads to a description of the system in terms of a generalized 
Gibbs ensemble~\cite{rigol07,cassidy2011,caux2013,lazarides14}. 

The main interests in the study of quantum statistical physics is therefore 
not only to see how a system equilibrates under the unitary evolution 
generated by its Hamiltonian, but also to investigate the nature and 
relaxation of a system driven out of equilibrium by a time-dependent 
Hamiltonian towards a nonequilibrium steady state (NESS). 
Due to the tremendous experimental 
progress~\cite{gring12,trotzky12,cheneau12,fausti11,rechtsman13,schreiber15}, 
a plethora of works has been carried out on periodically 
driven closed quantum systems~\cite{oka09,kitagawa10,lindner11,thakurathi13,
cayssol13,bastidas11,dasgupta15,privitera16,zhang17,choi17,russomanno12,
ponte15,mukherjee08,das10,alessio13,leon13,nag14,agarwala16,sharma14,
anirban_dutta14,russomanno15,russomanno16,sen16,bukov16,gritsev17,alessio14,
nandy17,utso18,ishii18,dumitrescu18,lazarides15,haldar17,bukov}. A time 
periodic Hamiltonian generates far richer possibilities for stabilizing 
a NESS with purely unitary dynamics, also rendering the possibility of 
exotic phases such as a Floquet time crystal~\cite{khemani16,else16}, Floquet 
Majoranas and other novel topological phases~\cite{kitagawa10,lindner11,
thakurathi13,cayssol13}. For Floquet systems which are integrable by a 
Jordan-Wigner transformation (from spin-1/2's to spinless fermions), the local 
observables eventually exhibit a steady state behavior which is described by 
a periodic Gibbs ensemble which is constructed via the entropy maximization 
principle by taking into account all the stroboscopically conserved 
quantities~\cite{lazarides14}. On the other hand, non-integrable systems
in the absence of disorder generally suffer from a heat death 
and eventually reaches an infinite temperature ensemble (ITE)~\cite{alessio14}.

In recent works, driving protocols that are not periodic functions of time 
have been considered~\cite{nandy17,utso18,ishii18,dumitrescu18,quelle17,ott84,
guarneri84,roosz16,martin17,nandy18}. For Jordan-Wigner integrable systems, it 
has been shown that any typical realization of random noise causes eventual 
heating to an ITE for all local observables. However, noise that is 
self-similar in time can eventually lead to an emergent steady state which is 
described by a geometric generalized Gibbs ensemble~\cite{nandy17}.
On the other hand, subjecting a disordered interacting spin chain to a 
quasiperiodic time-dependent Fibonacci drive typically leads to a long-lived 
glassy regime that eventually thermalizes to an ITE~\cite{dumitrescu18}. 

Motivated by the above considerations, we study in this paper an intermediate 
case between periodic and random driving of a Jordan-Wigner integrable quantum 
many-body system. Our system, although integrable, will be taken to be driven 
according to a quasiperiodic driving which follows the Fibonacci sequence. 
We ask whether such a driving protocol will cause heating to an ITE or 
saturation to a steady state for the local operators. Interestingly, we 
find that quasiperiodic driving leads to a a NESS in the high frequency limit. 
Furthermore, the time scale in which the system reaches a NESS is comparable 
to that of periodic driving and is therefore experimentally observable. 
This is in contrast to the scale-invariant situation in Ref.~\cite{nandy17}, 
where a NESS appears 
only at astronomically large times. We will further illustrate to what extent 
the generator of the quasiperiodic evolution can be reduced to an effective 
Hamiltonian whose spectrum in turn quantifies both the asymptotic value 
and the nature of the approach towards the NESS.

\begin{figure}[]
\centering
\subfigure[]{%
\includegraphics[width=.35\textwidth,height=.25\textwidth]{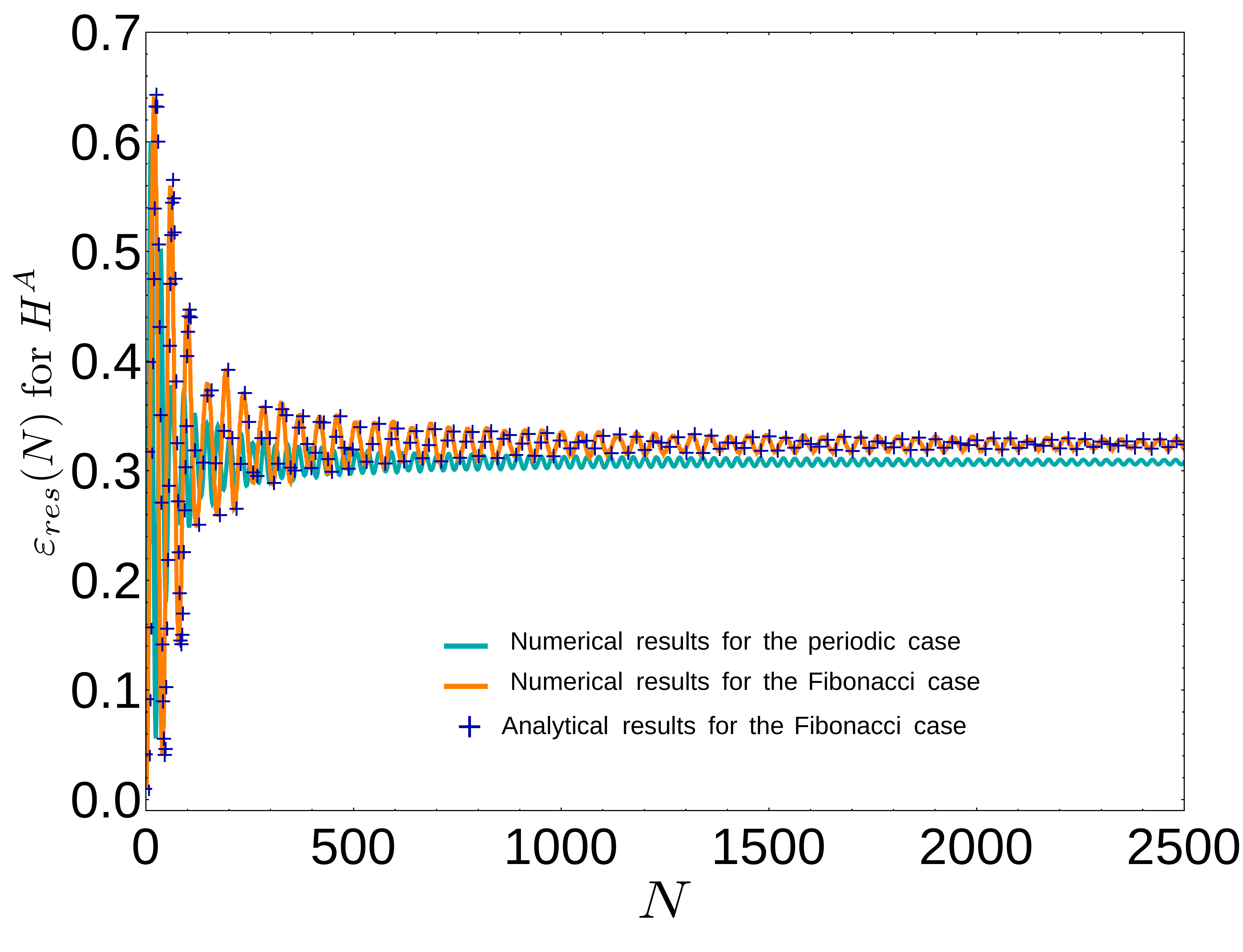}
\label{fig_ness}}
\quad
\subfigure[]{%
\includegraphics[width=.3\textwidth,height=.3\textwidth]{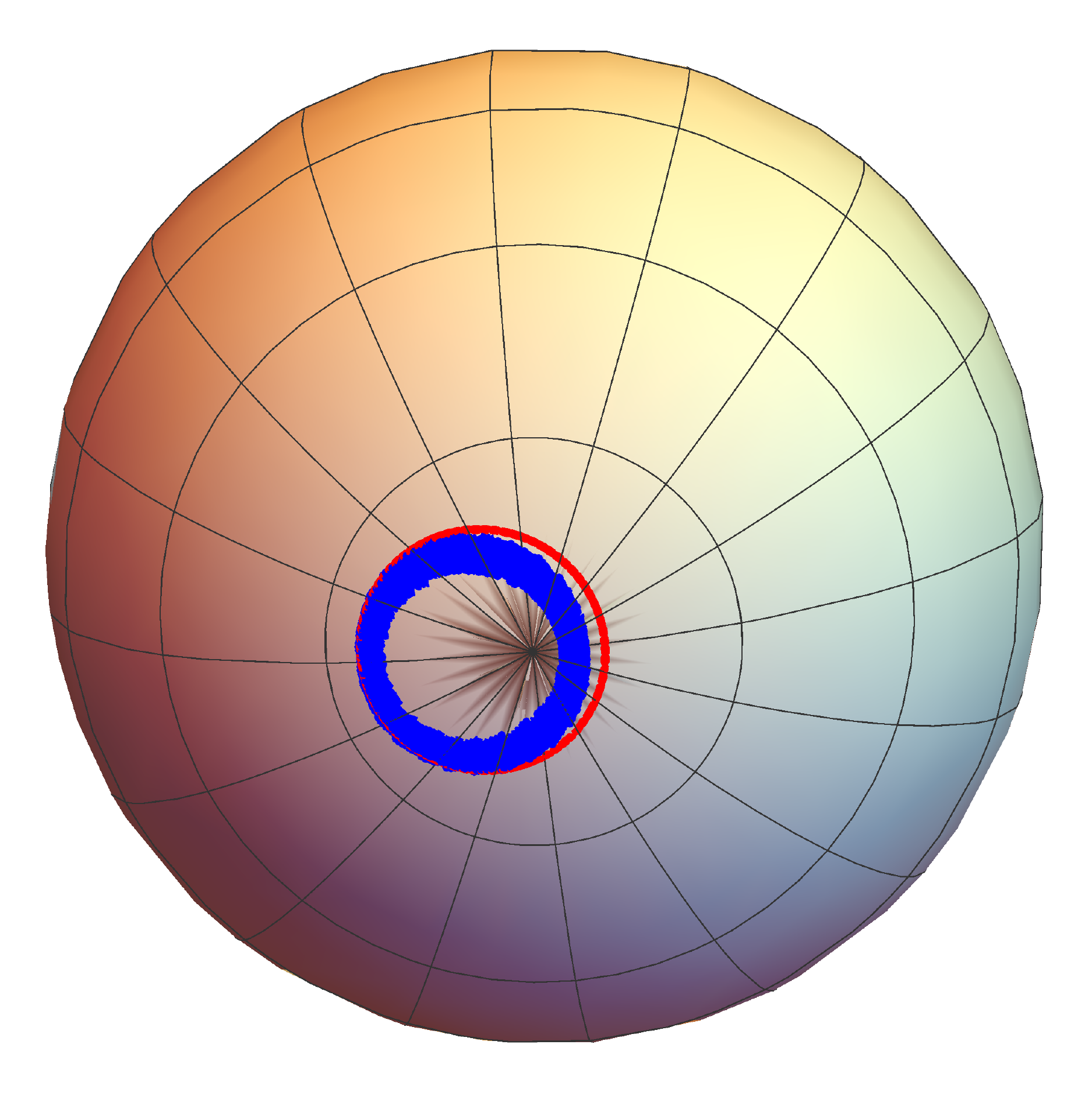}
\label{fig_bss}}
\quad
\subfigure[]{%
\includegraphics[width=.35\textwidth,height=4cm]{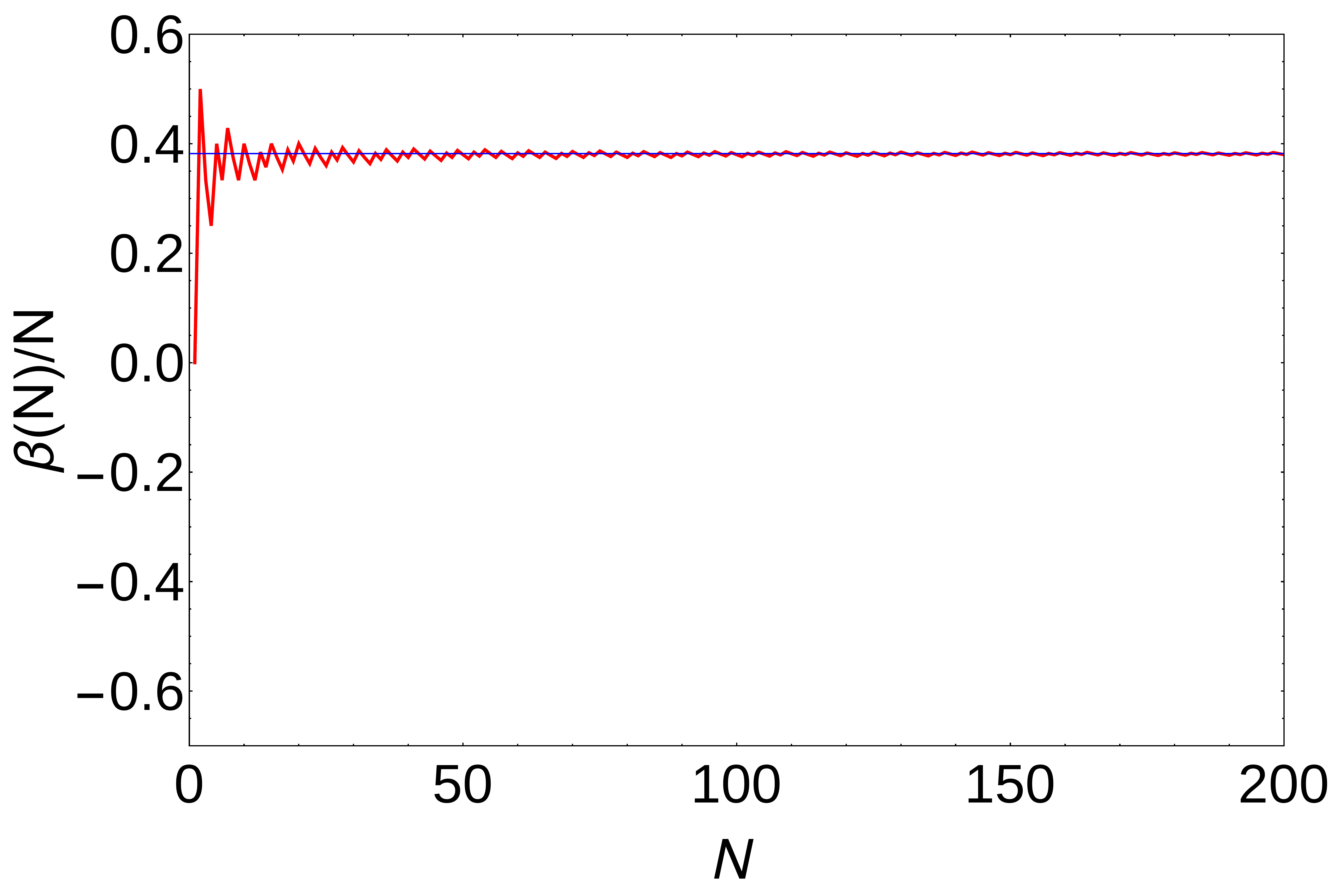}%
\begin{picture}(0,0)
\put(-130,20){\includegraphics[width=4.0cm,height=2.5cm]{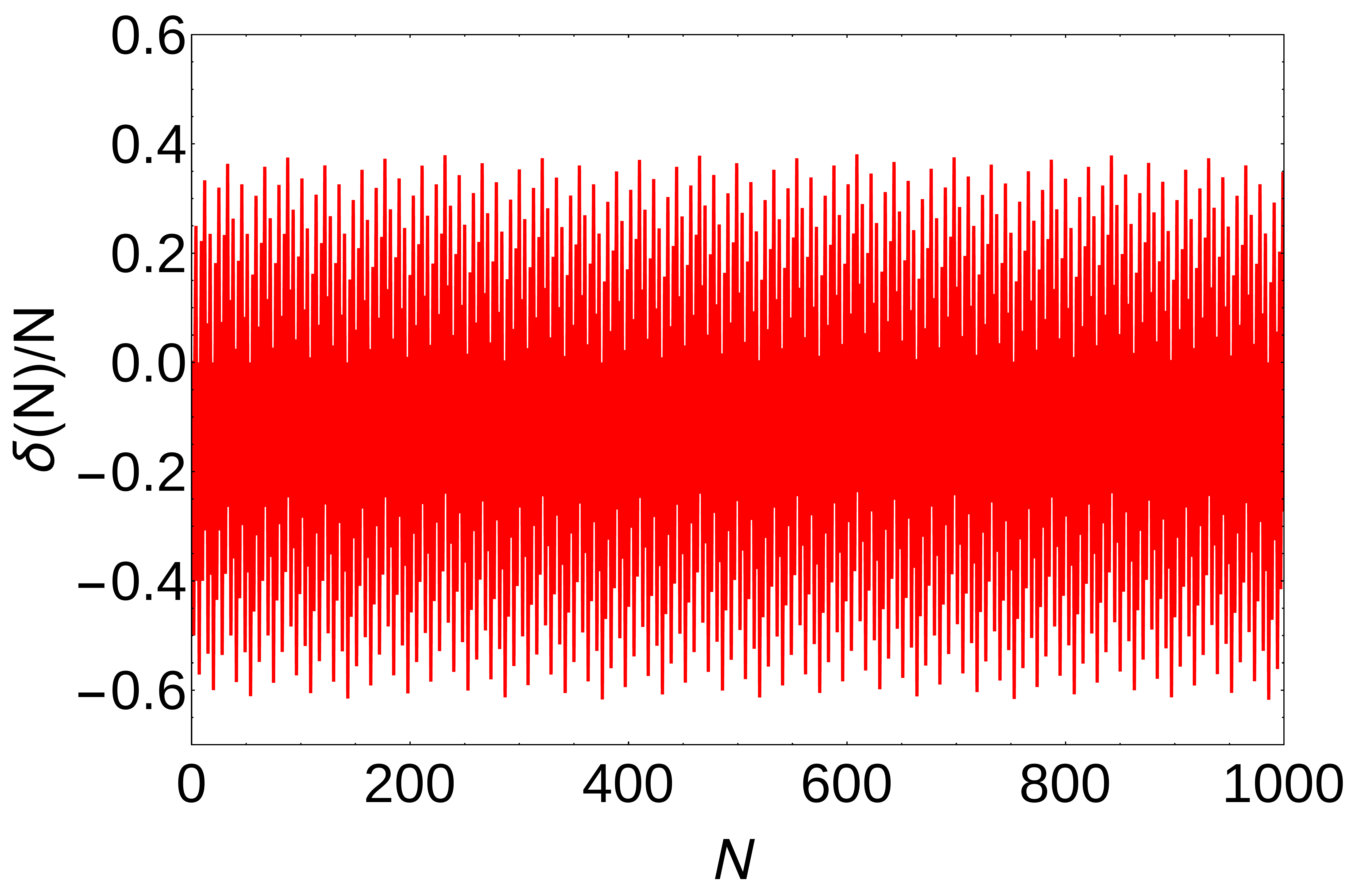}}
\end{picture}
\label{fig_ad}}
	
\caption{(a) Numerical results for the RE, $\sum_k [\bra{\psi_k(N)} H_k^{A}
\ket{\psi_k(N)}-$ $e^g_k(0)]/L$, vs $N$.
The cyan curve shows the RE for perfectly periodic driving with the protocol 
given in Eq.~\eqref{eq_prot}. The orange curve show the RE 
when $\ket{\psi_k(N)}$ is generated by the Fibonacci 
sequence. The plus signs indicate the approximate analytical results for the 
Fibonacci case using the unitary evolution in Eq.~\eqref{u_appx}, and they 
show an excellent fit to the numerical results.
(b) Trajectory $\{\theta_k(N),\phi_k(N)\}$ of the state $\ket{\psi_k(N)}$ 
in Eq.~\eqref{eq_staten} on the Bloch sphere as a function of $N$ up to 
$N=1000$ for a particular momentum mode $k= 159\pi/200$. 
In (a) and (b), we have chosen $L=500$, $\om =500$, and 
$h_A=1$ and $h_B=10$ in Eq.~\eqref{eq_prot}. The trajectory 
(in red) for the periodic dynamics (with the protocol in 
Eq.~\eqref{eq_prot}) forms a circle on the Bloch sphere. The 
trajectory for the Fibonacci driven sequence (in blue) fluctuates within the 
area bounded by two nearby and concentric circles lying on the Bloch sphere. 
(c) The figure shows that $\beta(N)/N$ quickly reaches a steady state value, 
equal to $0.382$ shown by the blue line, as $N$ becomes large. The inset shows 
that $\de (N)/N$ keeps fluctuating with the same amplitude even when 
$N$ becomes large.} \end{figure}

We consider the paradigmatic one-dimensional transverse field Ising model as 
an example of a Jordan-Wigner integrable system~\cite{lieb61,kogut79,suzuki13,
dutta15}. For each momentum mode, this is described by a $2 \times 2$ 
Hamiltonian~\cite{SM}
\begin{equation}\label{eq_ham}
H_k(t) ~=~ [h(t)-\cos k] ~\si_z ~+~ \sin k ~\si_x, \end{equation}
where the $\si$'s are Pauli matrices. 
We first consider a perfectly periodic driving with $H_k(t+\tau)=H_k(t)$, 
where the time period is $\tau=2T$, with a square pulse driving protocol of 
the form
\begin{eqnarray}\label{eq_prot}
H_k(t) ~=~ \begin{cases} H_k^A \hspace{3mm} \text{for} \hspace{3mm} 0 ~\leq~ 
t ~<~ T, \\
H_k^B \hspace{3mm} \text{for} \hspace{3mm} T ~\leq~ t ~<~ 2T, \end{cases}
\end{eqnarray}
where $H_k^A$ and $H_k^B$ are the momentum space Hamiltonian given in 
Eq.~\eqref{eq_ham} with transverse fields $h_A$ and $h_B$, respectively 
(see Ref.~\ct{SM} for details). For such a periodic protocol the system 
reaches a periodic steady state and the residual energy density (RE) reaches
a steady state value \cite{russomanno12,utso18,SM} (see the cyan 
curve in Fig.~\ref{fig_ness}). [We recall that the RE is defined as 
$\varepsilon_{res}(t)\equiv (1/L) \sum_{k} [e_k(t) - e_k^g(0)]$, where
$e_k(t)=\langle \psi_k(t) \rvert H_k(t)\lvert \psi_k(t) \rangle $ and 
$e_k^g(0)=\langle \psi_k(0) \rvert H_k(0) \lvert \psi_k(0) \rangle $, 
$\ket{\psi_k(t)}$ is the time-evolved state starting from the initial state 
$\ket{\psi_k(0)}$, $H_k(0)$ and $H_k(t)$ are the initial and instantaneous 
Hamiltonians of the system respectively, and $L$ is the system size].

We will now study the effect of a quasiperiodic driving protocol corresponding 
to a Fibonacci sequence of two distinct square wave pulses $A$ and $B$ (with 
Hamiltonians $H^A$ and $H^B$ respectively in Eq.~\eqref{eq_prot}), beginning 
as $ABAABABAABAAB \cdots$; we choose the first pulse to be $A$. We generate 
the Fibonacci sequence using the recursion relation
\begin{equation} V_n = V_{n-2}V_{n-1} \end{equation}
for $n\geq 2$ with two initial unitary matrices $V_0=U_B$ and $V_1 = U_A$; 
here $U_A$ and $U_B$ are evolution operators defined over a stroboscopic 
time $T$ for two different integrable Hamiltonians $H^A$ and $H^B$, such that
\begin{eqnarray}
U_A &=& e^{-iTH^A} ~\equiv~ e^A, \non \\
U_B &=& e^{-iTH^B} ~\equiv~ e^B.
\end{eqnarray}
We will measure the local observables after $N$ stroboscopic intervals, $t=NT$. 
The unitary operators for the first few values of $N$ are given by
\begin{eqnarray}
U(N=1) &=& e^A, \non \\
U(N=2) &=& e^B e^A \simeq e^{B + A + \frac{1}{2}[B,A]}, \label{eq_v2} \\
U(N=3) &=& e^A e^B e^A \simeq e^{A+B + A + \frac{1}{2}[B,A] + 
\frac{1}{2}[A,B]} \label{eq_v3}, \end{eqnarray}
and so on. We note that the last two approximations in Eqs.~\eqref{eq_v2} 
and \eqref{eq_v3} involve the multiplication of non-commuting matrices $e^A$
and $e^B$ and the application of the Baker-Campbell-Hausdorff formula retaining
only leading order terms in $1/\om$. The underlying assumption here is that 
the frequency $\om = 2\pi/T$ is much greater than the bandwidths of the two 
static Hamiltonians $H_k^A$ and $H_k^B$. For each $k$ mode, we can calculate 
the evolution operator $U_k(N)$ after $N$ stroboscopic intervals as (see 
Ref.~\cite{SM})
\begin{eqnarray}
U_k(N) &\simeq& e^{-iT\left( \al(N) H_k^A + \beta(N) H_k^B - i (T/2) \de 
(N) [H_k^A,H_k^B] \right)} \non \\
&\equiv& e^{-iNT H_k^{Fib}(N)}. \label{u_appx} \end{eqnarray}
Here
\begin{eqnarray} \beta(N) &=& 2N ~-~ \sum_{m = 1}^N \ga (m), \non \\
\al(N) &=& N ~-~ \beta(N), \non \\
\de (N) &=& \sum_{m=1}^N \left\{ \left[\ga (m) - 1\right](m - 1) - 
\flb{\frac{mG}{G+1}} \right\}, \non \\
\ga (m) &=& \fl{(m+1)G} ~-~ \fl{mG}, \end{eqnarray}
where $G = \left(\sqrt{5}+1\right)/2$ is the Golden ratio, and $\fl{x}$ 
denotes the largest integer $\le x$. We note that the function $\ga (m)$
is equal to either 1 or 2 for any positive integer $m$. 
We can now define an effective Hamiltonian $H_k^{Fib}(N)$ which is the 
generator of $U_k(N)$ as shown in Eq.~\eqref{u_appx},
\begin{eqnarray} H_k^{Fib}(N) &=& a_1\si_x + a_2\si_y + a_3\si_z, 
\end{eqnarray}
where the coefficients $a_i$ are given by
\begin{eqnarray}
a_1 &=& \sin k, \non \\
a_2 &=& \frac{\de (N)}{N} ~T\Delta h \sin k, \non \\
a_3 &=& h_A ~-~ \cos k ~+~ \frac{\beta(N)}{N}\Delta h, \end{eqnarray}
and $\Delta h = h_B - h_A$ is the amplitude difference of the two pulses. 
This effective Hamiltonian $H_k^{Fib}(N)$, in contrast to the Floquet 
Hamiltonian in the periodic case, depends on the stroboscopic time $N$; thus,
it yields time-dependent eigenvalues and eigenvectors which determine the 
behavior of the expectation values of local observables at all stroboscopic 
times. 

Before evaluating a local observable, we examine the dynamics of the time 
evolved state $\ket{\psi_k(N)}$ vs $N$ on the Bloch sphere for each $k$ 
mode. This state is numerically generated by acting with the Fibonacci 
evolution operator $U_k(N)$ on the initial state $\ket{\psi_k(0)}$ to yield
\begin{equation}\label{eq_traj}
\ket{\psi_k(N)}= U_k(N)\ket{\psi_k(0)} = \begin{bmatrix}
\cos\left(\theta_k(N)/2\right) \\
\sin\left(\theta_k(N)/2\right) e^{i\phi_k(N)} \\
\end{bmatrix}. \end{equation}
In Fig.~\ref{fig_bss}, we show the trajectory of $(\theta_k(N),\phi_k(N))$ 
for a particular $k$ mode on the Bloch sphere as it evolves with increasing 
$N$. We note that in contrast to the trajectory for the case of 
periodic driving shown by the red circle, the trajectory of the points
for the Fibonacci driving fluctuates but always
lies in the area bounded by two nearby and concentric circles lying on the 
Bloch sphere. This behavior can be understood by noting that although 
$\beta(N)/N$ quickly reaches a steady state value equal to $1-1/G \simeq 
0.382$ as $N$ becomes large, $\de (N)/N$ continues to fluctuate 
even for very large $N$; see Fig.~\ref{fig_ad} and its inset. The persistent 
fluctuations in $\de (N)/N$ prevents the trajectory from collapsing on to 
a single circle such as in the periodic case. The 
spread of the trajectory on the Bloch sphere is $k$-dependent and is directly
related to the amount of fluctuations of $\de (N)/N$. Reference~\cite{SM} 
provides an analytical derivation of $\beta(N)/N$ and the spread in 
$\de (N)/N$ which is found to lie between $1-1/G$ and $-1/G \simeq -0.618$.

\begin{figure}[]
\centering
\subfigure[]{%
\includegraphics[width=.45\textwidth,height=5cm]{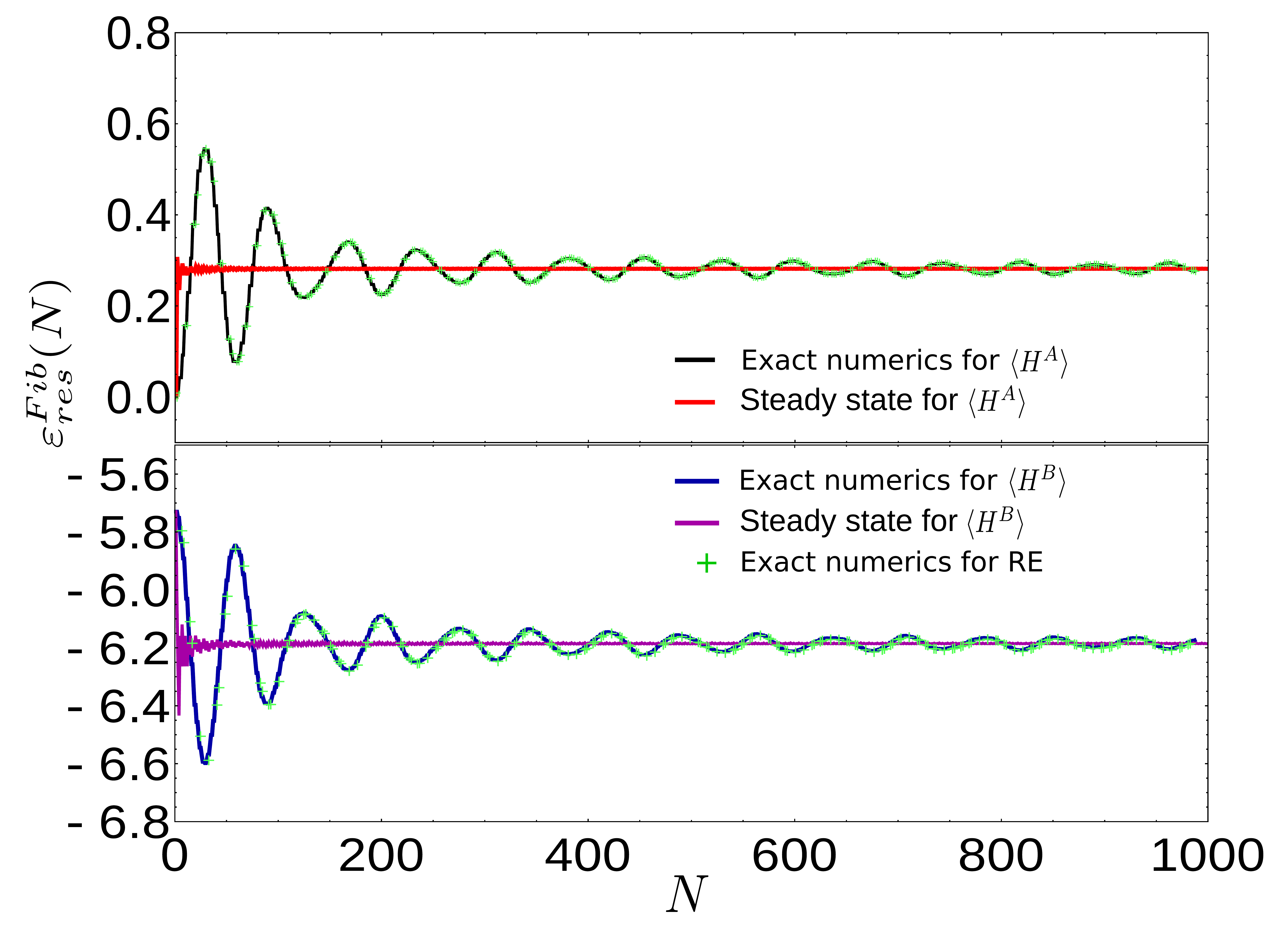}
\label{fig_eresab}}
\quad
\subfigure[]{%
\includegraphics[width=.41\textwidth,height=5cm]{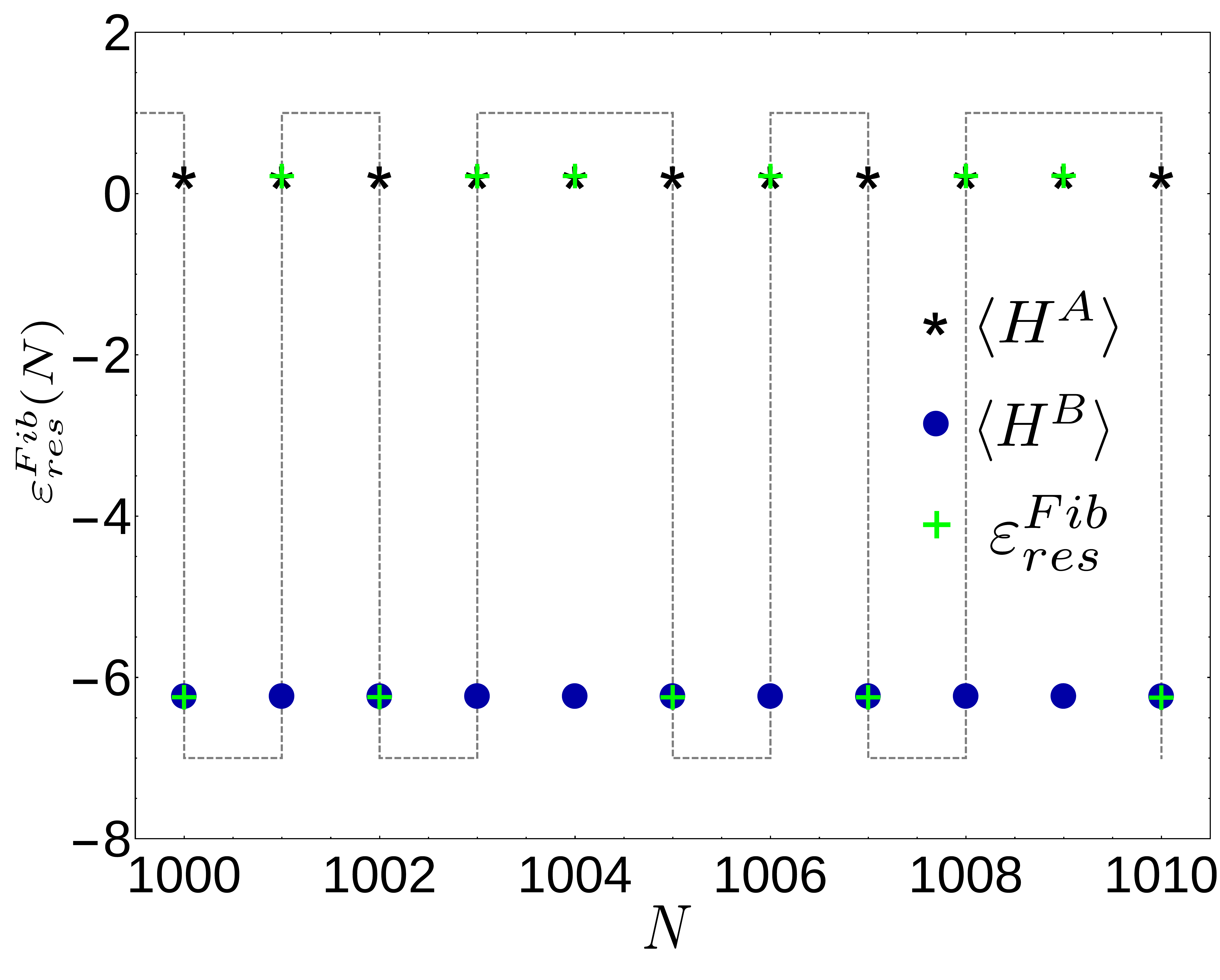}
\label{fig_ez}}
\caption{(a) $\langle H^{A}\rangle$ and $\langle H^{B}\rangle$ 
in Eq.~\eqref{eq_hab_exp} plotted vs $N$ for a 
Fibonacci protocol driven system with $h_A=1$, $h_B=10$, 
$\om = 500$, and $L=500$. The solid black (blue) lines show the numerically 
obtained results for $\langle H^{A}\rangle$ ($\langle H^{B}\rangle$), while 
the red (magenta) lines are the corresponding analytical steady state values. 
The green plus signed markers show how the RE (given in 
Eq.~\eqref{eq_redsplit}) is supported on the two quantities $\langle H^{A}
\rangle$ and $\langle H^{B}\rangle$. (b) This zoomed section of (a) 
focuses on the part between $N = 1000$ and $1010$. The black (blue) markers 
indicate the values of the $\langle H^{A}\rangle$($\langle H^{B}\rangle$). 
Following the green markers placed on top of the black (blue) markers, we 
observe that the steady state RE has support only on the upper (lower) branch 
when the system is viewed along the $A$ ($B$) sequence in the Fibonacci series 
shown by the gray dashed line.} \label{fig_resvn}
\end{figure}

Given the trajectory of the Fibonacci time-evolved state on the Bloch sphere, 
we are now ready to study whether the system 
attains a steady state asymptotically. To this end, we calculate the RE
in analogy with that of a perfectly periodic situation where the RE is given by 
the expectation value of the time-independent Hamiltonian $H_k(N)=H_k^A$ 
summed over all momenta modes. For the case of Fibonacci driving, we find
that the Hamiltonian is $N$ dependent and is given by
\begin{equation} H_k(N) ~=~ \left[\ga (N)-1\right]H_k^A ~+~ 
\left[2-\ga (N)\right] H_k^B. \label{eq_HN} \end{equation}
Since $\ga (N)$ is equal to either 1 or 2, $H_k(N)$ 
can either be $H_k^A$ or $H_k^B$ for each $N$. Then the RE is evaluated as 
$\varepsilon^{Fib}_{res}(N) = (1/L) \sum_k \left[\langle \psi_k(N) \rvert 
H_k(N) \lvert \psi_k(N) \rangle - e_k^g(0)\right]$. Using the high frequency 
approximation \eqref{u_appx}, the state after $N$ stroboscopic intervals can 
be written as 
\begin{equation}\label{eq_staten} 
\lvert \psi_k(N) \rangle = e^{-iNTH_k^{Fib}(N)}\ket{\psi_k(0)}. \end{equation}
Using the basis of eigenstates of $H_k^{Fib}(N)$, we can evaluate the RE 
in the high frequency limit for a thermodynamically large system with 
$L \to \infty$,
\begin{eqnarray}\label{eq_fibres}
\varepsilon_{res}^{Fib} (N) &=& \int\frac{dk}{2\pi} \{ \lvert c_k^{+}(N)
\rvert^2 H_k^{++}(N) + \lvert c_k^{-}(N) \rvert^2 H_k^{--}(N) \non \\
&& + ~(e^{iNT [ \mu_k^{+}(N)-\mu_k^{-}(N) ]} c_k^{+*}(N) c_k^{-}(N) 
H_k^{+-}(N) \non \\
&& ~~~~+~ \text{c.c.}) ~-~ e_k^g(0) \}, \end{eqnarray}
where $c_k^{\pm}(N)= \langle f_k^{\pm}(N)\rvert \psi_k(0) \rangle $, and 
$\ket{f_k^{\pm}}(N)$ are the eigenstates with eigenvalues $\mu_k^\pm(N)=
\sqrt{a_1^2+a_2^2+a_3^2}$ of the Hamiltonian $H_k^{Fib}(N)$. The 
matrix elements of $H_k(N)$ in this basis are given by
\begin{eqnarray}
H_k^{\pm\pm}(N)=\bra{f_k^{\pm}(N)}H_k(N)\ket{f_k^{\pm}(N)}. 
\label{eq_HN_element}\end{eqnarray}
\textit{In the limit of large $N$}, the off-diagonal terms containing 
$H_k^{+-}(N)$ and its complex conjugate in Eq.~\eqref{eq_fibres} oscillate 
rapidly and vanish on integrating over all the $k$ modes due to the 
Riemann-Lebesgue lemma, giving the steady state expression 
\begin{eqnarray} 
&~&\varepsilon^{Fib}_{res} \non \\
&=& \int \frac{dk}{2\pi} ~\{\vert c_k^+(N)\vert^2 H_k^{++}(N)
+ \vert c_k^-(N) \vert^2 H_k^{--}(N) 
 - e_k^g(0) \} \non \\
&=& \left[\ga (N)-1\right]\langle H^A\rangle ~+~ 
\left[2-\ga (N)\right]\langle H^B\rangle , 
\label{eq_redsplit}
\end{eqnarray}
where we have used Eqs.~\eqref{eq_HN} and \eqref{eq_HN_element} and 
the terms $\langle H^{A/B}\rangle$ are given by,
\begin{eqnarray}
\langle H^{A/B}\rangle=\int \frac{dk}{2\pi} \left\{\sum_{j=\pm}\vert c_k^j(N)
\vert^2\bra{f_k^{j}(N)}H_k^{A/B}\ket{f_k^{j}(N)}\right\}.
\label{eq_hab_exp} \end{eqnarray}
The quantities $\langle H^{A/B}\rangle$ can be obtained from the 
expectation values of $H^{A/B}$,
\begin{eqnarray}
\langle H^{A/B}(N)\rangle=\int \frac{dk}{2\pi} \left\{\bra{\psi_k(N)} 
H_k^{A/B} \ket{\psi_k(N)}-e_k^g(0)\right\}. \label{eq_hab} \end{eqnarray}
In taking the limit of large $N$, we drop the highly oscillating 
off-diagonal terms as they vanish upon integration over all the momentum
modes in the thermodynamic limit. Although the diagonal terms depend on $N$ 
through the basis vectors $\ket{f_k^{\pm}(N)}$, the quantities $\langle H^{A}
\rangle$ and $\langle H^{B}\rangle$ reach a steady state for a 
thermodynamically large system in the limit of large $N$ 
as shown in Fig.~\ref{fig_eresab}. The system reaches a steady state because 
the contributions of the fluctuating $\de (N)/N$ terms to $\langle H^{A/B}
\rangle$ vanish up on integration over the $k$ modes. Moreover, the steady 
state value of $\langle H^{A/B}\rangle$ also depends on $\beta(N)/N$ which, 
after some initial transients, settles to a value equal to $1-1/G$ and 
becomes independent of time.

Here, we would like to remark about the role of the $N$-dependence of 
$\gamma (N)$ in Eq.~\eqref{eq_redsplit}. In the case of a perfectly periodic 
driving, the steady state attains a constant value only when the system is 
observed at asymptotic stroboscopic instants $N$. There could of course be 
micromotion present in the system within a stroboscopic interval. If the 
system is observed at such intermediate times, it may not appear as steady. 
Similarly, in the
case of the Thue-Morse sequence~\cite{nandy17}, the steady state 
emerges only when it is observed at geometric intervals of $2^N$ and not at 
stroboscopic intervals $N$. In our case, the steady state state only attains 
a constant value when it is observed at the $A$ or $B$ stroboscopic instants 
of the Fibonacci sequence. Of course, it quasiperiodically oscillates between 
two different constant values (see Fig.~\ref{fig_ez}) if the system is instead
observed at each stroboscopic instant $N$. But, if we observe the system at the
time instances $A$ or $B$, then the corresponding steady state has only one 
constant value.

In summary, we have studied the behavior of a transverse Ising chain 
subjected to a Fibonacci driving protocol. For periodic driving, the evolution 
of each momentum mode on the Bloch sphere observed for a sufficiently long 
duration lies on a circle. In contrast, for the case of Fibonacci driving, 
we find in the high frequency limit that the evolving points lie within a small
area bounded by two concentric circles on the surface of the sphere; we have 
provided an explanation for this in terms of small but persistent fluctuations 
in the evolution operator. (It turns out that the axis of rotation of the 
circle changes after an astronomically large number of drives. Namely, 
the direction of the axis changes by $\ep$ after a number of drives of 
the order of $G^{\ep c /T^2}$, where $c$ is a number of order 1. For some
fixed values of $\ep$ and $c$, this is an enormously large number if $T$ is 
very small. Thus a change in the axis and therefore in the steady state 
would not be discernible within 
experimental time scales. This analysis is presented in Ref.~\cite{SM}).
Thus we have the interesting result that a thermodynamically large many-body 
system viewed stroboscopically reaches a different Fibonacci NESS and does not
heat up to an ITE in the limit of large $N$. Rather, when viewed after 
stroboscopic intervals $N$, the RE oscillates between two steady state values 
of the REs of the two Hamiltonians $H^A$ and $H^B$. These oscillations 
are quasiperiodic and exactly follow the Fibonacci sequence.
Whenever the sequence hits $A$ or $B$, the RE of the system 
follows the steady state RE calculated using $H^A$ or $H^B$. Thus, if the 
residual energy of the system is measured not after every stroboscopic 
interval, but either along the $A$'s in the Fibonacci sequence or along the 
$B$'s, it would appear that the steady state value of the RE is equal to the 
steady residual energy measured with respect to either $H^A$ or $H^B$, 
respectively. It is worth noting that the system has two accessible 
steady states between which the ones associated with $\langle H^B\rangle$ 
release energy and have a negative RE (see Fig.~\ref{fig_resvn}) compared to 
the initial state. This negative value of the RE occurs due to a greater 
population of those energy levels of $H^B$ which have lower energy than that 
of the initial ground state. This negative value can be tuned by varying the 
frequency $\om$ and the field $h_B$ with respect to $h_A$. In comparison, 
the RE in the perfectly periodic situation is always semi-positive. 
To conclude, in spite of the quasiperiodic nature of the driving, it is 
remarkable that the local properties of the system in the long-time 
limit manage to synchronize with the quasiperiodic drive and can eventually 
be described by a different nonequilibrium statistical ensemble.

We establish these findings by analytically deriving an effective Hamiltonian 
$H_k^{Fib}(N)$, which is $N$-dependent unlike the periodic Floquet scenario
and can nearly exactly simulate the dynamics of the system in the 
high frequency limit (see Fig.~\ref{fig_ness}). 
The time-dependent spectrum of $H_k^{Fib}(N)$ can effectively provide a 
microscopic understanding of the nature of evolution towards a steady state 
as $N$ becomes large.

We would like to conclude by highlighting that our work interestingly shows 
the emergence of a steady state behavior albeit only at high frequencies. 
The emergence of such a steady state and its form (namely, an annular spread 
of the eigenstates on the Bloch sphere) are not {\it a priori} obvious. 
The uniqueness of the steady state lies in the fact that when 
the system is observed perfectly periodically, the steady state value 
oscillates quasiperiodically following the Fibonacci sequence whereas, if it is
observed at Fibonacci instances, the system oscillates periodically between
two constant values. Furthermore, this emergent behavior is well explained
through an analytical framework devised using the high frequency approximation
which is in complete agreement with numerical simulations. The analytical
results allow us to explore the behavior of the system in the large $N$ limit, 
where numerical errors due to matrix multiplications eventually
creep in. We note that if the frequency is not high, no such steady state 
exists, and the system can exhibit a rich variety of long-time behaviors 
depending on the values of the driving parameters~\cite{nandy18}. The role of 
interactions and disorder and the eventual heating up to an ITE has been
investigated in Ref.~\cite{dumitrescu18}.

As long as the driving sequence is of the Fibonacci type,
the fact that the system eventually reaches a steady state is not restricted 
to the square pulse nature of the driving protocol,
though the steady state value of the RE may depend on the strength of 
the driving involved. The analytic evaluation of the RE assumes a 
knowledge of the binary non-commuting unitary evolution operators 
over a complete stroboscopic period whose generators 
are Jordan-Wigner integrable and are devoid of local disorder. Thus,
the same results are expected to hold for higher-dimensional systems as well. 
We note that the binary aperiodic situation comprising a 
$\de$-function kicking protocol has already been experimentally realized for 
a single rotor~\cite{sarkar17}; similar experimental studies for 
our quasiperiodically driven situation are indeed possible. 

We thank Sourav Nandy and Arnab Sen for many stimulating discussions. 
A.D. acknowledges SERB, DST, India and D.S. thanks DST, India for 
Project No. SR/S2/JCB-44/2010 for financial support.

\newpage

\begin{center}
\textbf{\large Supplemental Material on ``Fibonacci steady states in a driven 
integrable quantum system"} \\
\vspace{0.5cm}
\vspace{0.2cm}
\end{center}
\setcounter{equation}{0}
\setcounter{figure}{0}
\setcounter{table}{0}
\setcounter{page}{1}
\makeatletter
\renewcommand{\theequation}{S\arabic{equation}}
\renewcommand{\thefigure}{S\arabic{figure}}
\renewcommand{\bibnumfmt}[1]{[S#1]}
\renewcommand{\@cite}[1]{[S#1]}

\section{The model and the steady state for periodic driving}

We consider the one-dimensional transverse field Ising model as an example of 
a Jordan-Wigner integrable system [S1-S4]
This is described by the Hamiltonian
\begin{equation} \label{eq_model}
H(t) ~=~ - ~\sum_{n=1}^{L} ~\tau_n^{x}\tau_{n+1}^{x} ~-~ h(t) ~\sum_{n=1}^{L}
~\tau_{n}^{z}, \end{equation}
where $h$ is the time-dependent transverse field and $\tau_{n}^{a}$ $\{a=x,y,
z\}$ are the Pauli spin matrices at the $n^{th}$ site. 
Following a Jordan-Wigner transformation from spin-1/2's to spinless fermion 
operators at each site, the Hamiltonian gets decoupled 
into two-level systems for pairs of Fourier modes with momenta $\pm k$ (where 
$k$ lies in the range $[0,\pi]$), such that 
\begin{eqnarray} H(t) &=& \int_{0}^{\pi} \frac{dk}{2\pi} ~\left(
\begin{array}{cc}
c_k^\da & c_{-k} \end{array} \right) ~H_k(t) \left( \begin{array}{c}
c_k \\
c_{-k}^\da \end{array} \right), \non \\
H_k(t) &=& [h(t) ~-~ \cos k] ~\si_z ~+~ \sin k ~\si_x,
\label{eq_ham2} \end{eqnarray}
where the $\si$'s are again Pauli matrices. Here we have applied antiperiodic 
boundary conditions for even $L$ so that $k=2m\pi/L$ with $m=-(L-1)/2, \cdots, 
-1/2, 1/2, \cdots, (L-1)/2$. 
 
We first consider perfectly periodic driving with $H_k(t+\tau)=H_k(t)$, 
where the time period is $\tau=2T$, with a square pulse driving protocol of 
the form
\begin{eqnarray}\label{eq_prot2}
H_k(t) ~=~ \begin{cases} H_k^A \hspace{3mm} \text{for} \hspace{3mm} 0 ~\leq~ 
t ~<~ T, \\
H_k^B \hspace{3mm} \text{for} \hspace{3mm} T ~\leq~ t ~<~ 2T, \end{cases}
\end{eqnarray}
where $H_k^A$ and $H_k^B$ are the momentum space Hamiltonian given in 
Eq.~\eqref{eq_ham2} with transverse fields $h_A$ and $h_B$, respectively. For 
such a periodic protocol the system reaches a periodic steady state 
and the residual energy density (RE) reaches a steady state 
value~\cite{russomanno12} (see the cyan curve in Fig.~1(a) of the main text).
We recall here that the RE is defined as the difference between the
energy density of the time evolved state determined by the
expectation value of the instantaneous Hamiltonian and the energy density of
the initial state of the system. Namely,
$\varepsilon_{res}(t)\equiv (1/L) \sum_{k} [e_k(t) - e_k^g(0)]$, where
$e_k(t)=\langle \psi_k(t) \rvert H_k(t)\lvert \psi_k(t) \rangle $ and 
$e_k^g(0)=\langle \psi_k(0) \rvert H_k(0) \lvert \psi_k(0) \rangle $, 
$\ket{\psi_k(t)}$ is the time evolved state starting from the initial
state $\ket{\psi_k(0)}$, and $H_k(0)$ and $H_k(t)$ are the initial and 
instantaneous Hamiltonians of the system respectively.

We choose the ground state of the Hamiltonian $H_k^A$ to be the initial 
state $|\psi_k(0)\rangle$ for each $k$ mode. Using the Floquet formalism, we
can define a Floquet evolution operator $\mathcal{F}_k(\tau) = 
\mathcal{T}$ $\exp \left[ -i \int_{0}^{\tau}H_k(t) dt \right]$, where 
$\mathcal{T}$ denotes time ordering. We now recall that the solutions of the 
Schr\"{o}dinger equation for a time periodic Hamiltonian can be written as 
$\lvert \psi_k^{j}(t) \rangle = \exp(-i \ep_k^{j} \tau) \lvert \phi_k^{j}
(t) \rangle$, where the states $\lvert \phi_k^{j}(t) \rangle$ satisfying the
condition $\lvert \phi_k^{j}(t+\tau) \rangle =\lvert \phi_k^{j}(t) \rangle$ 
are called Floquet modes and the real quantities $\ep_k^{j}$ are 
known as Floquet quasienergies. For our square pulse protocol, the Floquet 
evolution operator in a single interval of $\tau$ can be written in the form 
\begin{eqnarray}\label{eq_perev}
\mathcal{F}_k(\tau) ~=~ e^{-iTH_k^B} ~e^{-iTH_k^A}, \end{eqnarray}
and is generated by two piecewise constant Hamiltonians $H_k^A$ and $H_k^B$. 
For a periodic drive of a system reducible to decoupled two-level 
systems, Eq.~\eqref{eq_perev} can be calculated analytically.
We now note that in the high-frequency limit (i.e., small $\tau$), 
Eq.~\eqref{eq_perev} can be approximated, using the Baker-Campbell-Hausdorff
(BCH) formula $\exp (C) \exp (D) = \exp (C + D + (1/2)[C,D] + \cdots)$, by
\begin{equation}
\mathcal{F}_k(\tau) ~=~ \exp [-i (\tau/2) (H_k^A+H_k^B + \tau \Delta_h
\sin k ~\si_y)], \end{equation}
where $\Delta_h = (h_B - h_A)/2$.
We note that this truncation scheme is similar to the Magnus expansion.

To establish that a periodically steady behavior [S5,S6]
exists for our Jordan-Wigner integrable system under the periodic protocol in 
Eq.~\eqref{eq_prot2}, we will now evaluate the RE of the system. For the 
periodically driven case in Eq.~\eqref{eq_prot2} with $H_k(N\tau)= H_k(0)=
H_k^A$, we have, after a time $t=N\tau$ made up of $N$ stroboscopic intervals, 
$\lvert \psi_k(N) \rangle = \sum_{j=\pm} r_k^{j} e^{-i \ep_k^{j}N \tau} \lvert 
\phi_k^{j}(t) \rangle$, where $r_k^{\pm}= \langle \phi_k^{\pm} \rvert 
\psi_k(0) \rangle $. In the thermodynamic limit $L \to \infty$, we obtain
\begin{equation}
\varepsilon_{res}(N) ~=~ \int\frac{dk}{2\pi} ~\biggl[ ~\sum_{\al = \pm} ~| 
r_k^{\al} |^2 \langle \phi_k^{\al} \rvert H_k^A\lvert \phi_k^{\al} \rangle 
- e_k^g(0) ~+~ \sum_{\substack{ \al,\beta = \pm \\ \al \neq \beta}} 
r_k^{\al \ast} r_k^{\beta} e^{i(\ep_k^\al - \ep_k^\beta ) N\tau} \langle 
\phi_k^{\al} \rvert H_k^A \lvert \phi_k^{\beta} \rangle ~\biggr]. 
\label{eq_eresp} \end{equation}
In the limit $N \to \infty$, the rapidly oscillating off-diagonal terms in 
Eq.~\eqref{eq_eresp} will vanish upon integration over all $k$ modes leading 
to a steady state expression for $\varepsilon_{res}(N)$~\cite{russomanno12} 
(see the cyan curve in Fig.~1 (a) of the main text). 

\section{Calculation of evolution operator $U(N)$ for Fibonacci driving}

We now present a detailed calculation of the stroboscopic evolution operator 
$U(N)$ for a Fibonacci sequence of driving generated by two type of pulses 
$A$ and $B$ with Hamiltonians $H^A$ 
and $H^B$, respectively. (All the discussion here will refer to a particular 
$k$ mode, and we will not show the label $k$ explicitly). Reading from left 
(earliest time) to right (latest time), the Fibonacci sequence is given by 
\begin{equation}\label{eq_sequence_ab}
ABAABABAABAABABAABABA \cdots. \end{equation}
The stroboscopic time evolution operators after the first few stroboscopic 
intervals $N$ are given by
\begin{eqnarray}
U(N=1) &=& e^{-i H^A T} \equiv e^A, \label{eq_n1} \\ 
U(N=2) &=& e^{-i H^B T}e^{-i H^A T} \equiv e^B e^A \approx e^{B+A+
\frac{1}{2}[B,A]}, \label{eq_n2} \\
U(N=3) &=& e^A U(N=2) \approx e^{A+B+A+\frac{1}{2}[B,A]+ \frac{1}{2}[A,B]}, 
\label{eq_n3} \\ 
U(N=4) &=& e^A U(N=3) \approx e^{A+A+B+A+\frac{1}{2}[B,A]+ \frac{1}{2}[A,B]+
\frac{1}{2}[A,B]}, \label{eq_n4} \\ 
U(N=5) &=& e^B U(N=4) \approx e^{B+A+A+B+A+\frac{1}{2}[B,A]+ \frac{1}{2}[A,B]+
\frac{1}{2}[A,B]+\frac{3}{2}[B,A]}, \label{eq_n5} \\
U(N=6)&=& e^A U(N=5) \approx e^{A+B+A+A+B+A+\frac{1}{2}[B,A]+ \frac{1}{2}[A,B]
+\frac{1}{2}[A,B]+\frac{3}{2}[B,A]+[A,B]}, \label{eq_n6} \end{eqnarray}
and so on. In Eqs.~\eqref{eq_n2} to \eqref{eq_n6}, we have used the BCH 
formula and considered only terms to leading order in the time period $T$ 
(i.e., up to $T^2$ to keep only the first order commutator of $A$ and $B$);
this is valid for small values of $T$. 
 
\begin{table}[h]
\centering
\caption{Stroboscopic Fibonacci sequence at Fibonacci steps ($N=F_n$)}
\label{table1}
\begin{tabular}{|p{6cm}|p{4cm}|p{4cm}|}
\hline
Fibonacci sequence of $A$'s and $B$'s & Total number of stroboscopic intervals 
$N$ & Number of $A$'s ($X$) and $B$'s ($Y$) \\
\hline
A & 1 ($F_2$) & $X=1$ $(F_1), Y=0$ $(F_0)$ \\
AB& 2 ($F_3$) & $X=1$ $(F_2), Y=1$ $(F_1)$ \\
ABA & 3 ($F_4$) & $X=2$ $(F_3), Y=1$ $(F_2)$ \\
ABAAB & 5 ($F_5$) & $X=3$ $(F_4), Y=2$ $(F_3)$ \\
ABAABABA & 8 ($F_6$) & $X=5$ $(F_5), Y=3$ $(F_4)$ \\
ABAABABAABAAB & 13 ($F_7$) & $X=8$ $(F_6), Y=5$ $(F_5)$ \\
ABAABABAABAABABAABABA & 21 ($F_8$) & $X=13$ $(F_7), Y=8$ $(F_6)$ \\
\vdots&\vdots&\vdots\\
\hline
\end{tabular}
\end{table}

To determine the evolution operator $U(N)$ for arbitrary $N$, let us first 
consider the case where $N$ is equal to a Fibonacci number $F_n$, i.e., at the 
Fibonacci steps as shown in Table~\ref{table1}. In the Fibonacci sequence with 
$N=F_n$, the number of $A$'s and $B$'s present are given by $F_{n-1}$ and 
$F_{n-2}$, respectively. This follows from the definition of the Fibonacci 
numbers
\beq F_{n+1} ~=~ F_n ~+~ F_{n-1}, \label{fib1} \eeq
with the first few Fibonacci numbers being given by $F_0 = 0$, $F_1 = 1$,
$F_2 = 1$, $F_3 = 2$, and $F_4 = 3$. Let $G=(\sqrt{5}+1)/2 \simeq
1.618$ denote the Golden ratio; we will repeatedly use the identity $G^2 = 
G + 1$ below. We can show from Eq.~\eqref{fib1} and the values of $F_0$ and
$F_1$ that
\beq F_n ~=~ \frac{1}{\sqrt{5}} ~\left[ G^n ~-~ \left(- \frac{1}{G} \right)^n
\right]. \label{fn} \eeq
For large $n$, we have $F_n \simeq G^n/\sqrt{5}$ and therefore $F_n/F_{n-1}
\approx G$. For large values of $N=F_n$, the number of $A$'s and $B$'s present
in the expression for $U(N)$ are therefore given by
\begin{equation}
X ~\approx~ \frac{NG}{1+G} \hspace{0.3cm} \text{and} \hspace{0.3cm} Y ~\approx~
\frac{N}{1+G}, \end{equation}
respectively, such that $X+Y=N$ and $X/Y\approx G$.

\begin{figure}[]
\centering
\includegraphics[width=.48\textwidth,height=.42\textwidth]{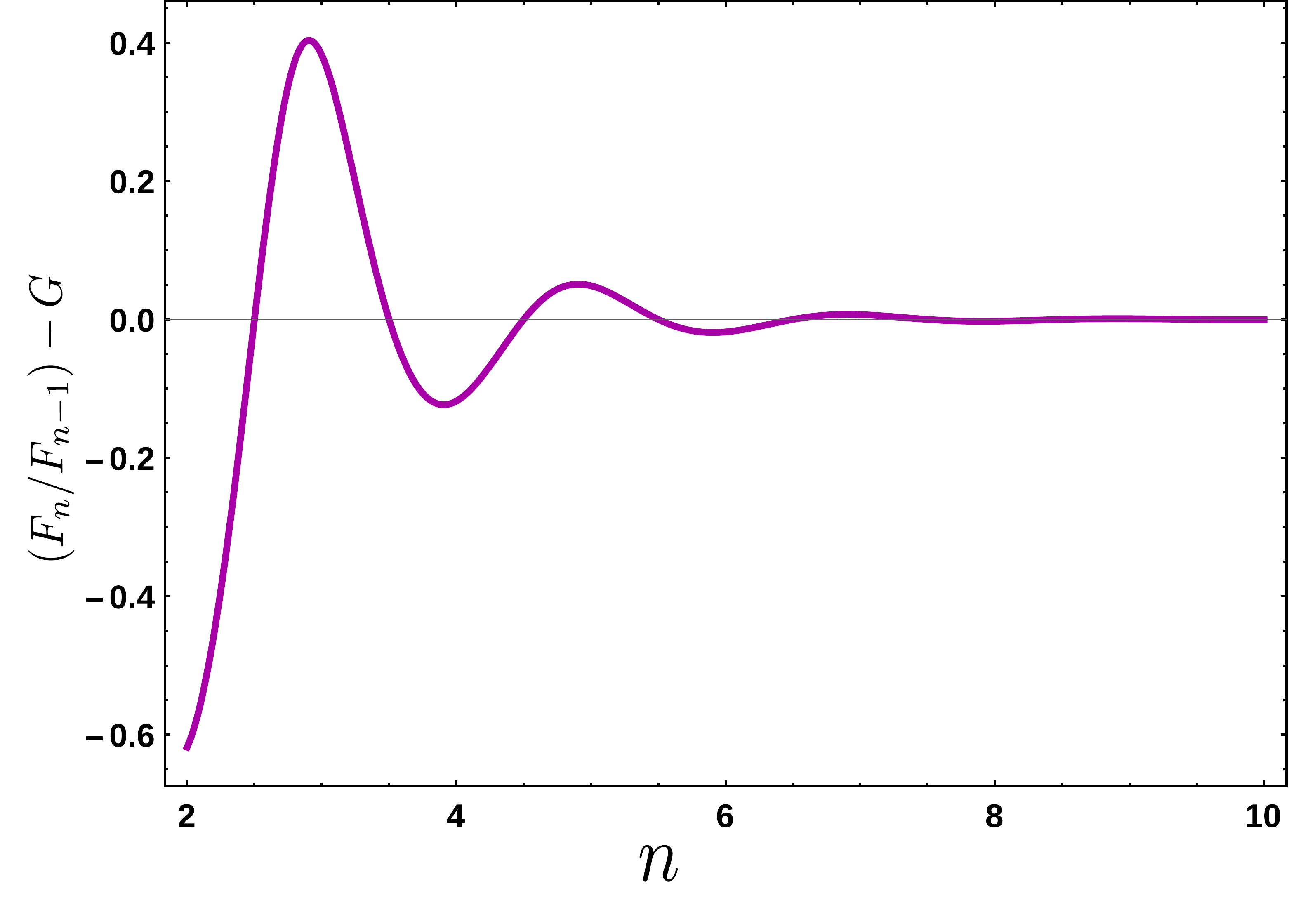}
\caption{Figure showing that the ratio $F_n/F_{n-1}$ approaches $G$ very
quickly as $n$ increases.} \label{fig_goldenr} \end{figure}

To calculate the number of $A$'s and $B$'s for an arbitrary value of $N$,
let us consider the function
\begin{equation}
\gamma(m)= \fl{(m+1)G} - \fl{mG}, \end{equation}
where $\fl{x}$ denotes the largest integer $\le x$. The function $\gamma (m)$ 
is equal to either 1 or 2 for any positive integer $m$. Therefore, the 
function $\gamma(m)-1$ is a Boolean function; interestingly, it generates the 
Fibonacci sequence of $1$'s and $0$'s as
\begin{equation}
101101011011010110101 \cdots \end{equation} 
for positive integer values of $m$. If we take 1 to correspond to $A$ and 0 
to correspond to $B$, the above sequence is exactly the same as the sequence 
given in Eq.~\eqref{eq_sequence_ab} and is represented by the function
\begin{equation}
S(N)=\left[\gamma(N)-1\right]A+\left[2-\gamma(N)\right]B. \end{equation}
For a particular stroboscopic instant $N$, if the function $\gamma(N)$ is 1 
(2), it implies that a $B$-pulse ($A$-pulse) is present. Hence, the total 
number of $B$ pulses in the sequence up to $N$ stroboscopic intervals is given 
by
\begin{equation}
\beta(N)= \sum_{m=1}^{N} \left\{ 2-\gamma(m) \right\},
\label{eq_alphan} \end{equation}
and the total number of $A$ pulses in the sequence is given by
\begin{equation}
\alpha(N)=\sum_{m=1}^{N} \left\{\gamma(m)-1\right\}=N-\beta(N).
\label{eq_betan} \end{equation}
Now, in order to calculate $U(N)$ in Eqs.~\eqref{eq_n2} to \eqref{eq_n6}, it 
is necessary to count the number of commutators $[A,B]$ or $[B,A]$. It is easy 
to see that depending on whether the pulse $A$ or $B$ is present at the 
$m^{th}$ stroboscopic instant (where $m\le N$), the number of commutators 
$[A,B]$ or $[B,A]$ is only given by the number of $B$ or $A$ pulses present in 
the sequence up to the $(m-1)^{th}$ stroboscopic instant. The function 
$\fl{(mG)/(1+G)}$ gives the number of $A$ pulses present in the sequence 
before the $m^{th}$ $B$ pulse and $\fl{(m)/(1+G)}$ gives the number of $B$ 
pulses present before the $m^{th}$ $A$ pulse, as illustrated in 
Table~\ref{table2}.

\begin{table}[h]
\centering
\caption{Illustration of the numbers of $[A,B]$'s and $[B,A]$'s.}
\label{table2}
\begin{tabular}{|p{2cm}|p{0.4cm}p{0.4cm}p{0.4cm}p{0.4cm}p{0.4cm}p{0.4cm}p{0.4cm}p{0.4cm}p{0.4cm}p{0.4cm}p{0.4cm}p{0.4cm}p{0.4cm}p{0.4cm}p{0.4cm}p{0.4cm}p{0.4cm}p{0.4cm}p{0.4cm}p{0.4cm}p{0.4cm}p{0.4cm}|}
\hline
$N=$ &1&2&3&4&5&6&7&8&9&10&11&12&13&14&15&16&17&18&19&20&21&$\cdots$ \\
\hline
$S(N)=$ &A&B&A&A&B&A&B&A&A&B&A&A&B&A&B&A&A&B&A&B&A&$\cdots$ \\
$\fll{\frac{m}{1+G}}=$ &0&0&1&1&1&2&2&3&3&3&4&4&4&5&5&6&6&6&7&7&8&$\cdots$ \\
$\fll{\frac{mG}{1+G}}=$ &0&1&1&2&3&3&4&4&5&6&6&7&8&8&9&9&10&11&11&12&12&
$\cdots$ \\
\hline
\end{tabular}
\end{table}

Now, we can calculate the number of commutators $[A,B]$ present in the 
evolution operators $U(N)$; this is given by
\begin{equation}
\delta(N) = \sum_{m=1}^{N} \left[ \left\{\gamma(m) -1 \right\}\flb{
\frac{m}{1+G}} - \left\{2-\gamma(m)\right\}\flb{\frac{mG}{1+G}}\right],
\end{equation}
where the first term comes from counting all the commutators $[A,B]$ due to 
the positions of $A$ pulses and and the second term comes from counting all 
the commutators $[B,A]$ due to the positions of $B$ pulses. Using the relation 
$\fl{m/(1+G)}+\fl{mG/(1+G)}=m-1$, one can easily simplify the above expression 
as
\begin{equation}
\delta(N)=\sum_{m=1}^{N} \left\{ (\gamma(m)-1)(m-1)-\flb{\frac{mG}{1+G}}
\right\}. \label{eq_deltan} \end{equation}
Using Eqs.~\eqref{eq_alphan}, \eqref{eq_betan} and \eqref{eq_deltan}, the 
stroboscopic evolution operator $U(N)$ for an arbitrary instant $N$ in the 
high frequency approximation is thus found to be
\begin{eqnarray}
U(N) &=& \exp \left\{ \alpha(N) A + \beta(N) B +\frac{\delta(N)}{2} 
[A,B] \right\} \label{un} \\ 
&=& \exp \left\{-i T \alpha(N) H^A -iT\beta(N) H^B -\frac{T^2\delta(N)}{2} 
[H^A,H^B] \right\}. \end{eqnarray}

\section{Calculation of the spread in the values of $\delta (N)/N$}

We will now present more explicit expressions for $\al (N), ~\beta (N)$ and 
$\de (N)$ as functions of $N$; this will enable us to precisely calculate
the spread in the values of $\de (N)/N$. We consider the expression in 
Eq.~\eqref{un} which we get after $N$ stroboscopic drives. Given two matrices 
$U (N)$ and $U (M)$, we find that
\beq U (P) ~\equiv~ U (M) ~U (N) \eeq
will have coefficients $\al (P), ~\beta (P), ~\de (P)$ given by
\bea \al (P) &=& \al (M) ~+~ \al (N), ~~~\beta (P) ~=~ \beta (M) ~+~ 
\beta (N), \non \\
{\rm and}~~~ \de (P) &=& \de (M) ~+~ \de (N) ~+~ \al (M) \beta (N) ~-~ 
\beta (M) \al (N), \label{abd} \eea
where we have used the BCH formula up to first order in the commutators.

The first few $U (N)$'s are given in Eqs.~\eqref{eq_n1} - \eqref{eq_n6}.
This tells us that
\bea \al (1) &=& 1, ~~~\beta (1) ~=~ 0, ~~~\de (1) ~=~ 0, \non \\
\al(2)&=& 1, ~~~\beta(2) ~=~ 1, ~~~\de (2) ~=~ -1, \non \\
\al(3)&=& 2, ~~~\beta(3)~=~ 1, ~~~\de (3) ~=~ 0, \non \\
\al(4)&=& 3, ~~~\beta(4)~=~ 1, ~~~\de (4) ~=~ 1, \label{abdn} \eea
and so on.

We now consider the case where $N=F_{n+1}$ is a Fibonacci number. Let us define
\beq V_n ~\equiv~ U (F_{n+1}). \label{uvn} \eeq
We then have the recursion relation
\beq V_n ~=~ V_{n-2}~ V_{n-1} \label{vn} \eeq
for $n \ge 2$, where $V_0 = e^B$ and $V_1 = e^A$. Using Eq.~\eqref{abd}, we 
find the recursion relations,
\bea \al (F_{n+1}) &=& \al (F_{n-1}) ~+~ \al (F_n), \non \\
\beta (F_{n+1}) &=& \beta (F_{n-1}) ~+~ \beta (F_n), \non \\
\de (F_{n+1}) &=& \de (F_{n-1}) ~+~ \de (F_n) ~+~ \al (F_{n-1}) ~\beta (F_n) ~
-~ \beta (F_{n-1}) ~\al (F_n). \label{abdn2} \eea
The first two equations in Eqs.~\eqref{abdn2} along with the values of $\al 
(1), ~\al (2), ~\beta (1), ~\beta (2)$ imply that
\beq \al (F_{n+1}) ~=~ F_n ~~~{\rm and}~~~ \beta (F_{n+1}) ~=~ F_{n-1}. 
\label{abn} \eeq
We then find that the last two terms in the third equation in 
Eqs.~\eqref{abdn2} give
\bea \al (F_{n-1}) ~\beta (F_n) ~ -~ \beta (F_{n-1}) ~\al (F_n) &=&
F_{n-2} ~F_{n-2} ~-~ F_{n-3} ~F_{n-1} \non \\
&=& - ~(-1)^n, \label{iden} \eea
where we have used Eq.~\eqref{fn} to derive the second equation from the first 
in Eq.~\eqref{iden}. The third equation in Eqs.~\eqref{abdn2} then takes the 
form
\beq \de (F_{n+1}) ~=~ \de (F_{n-1}) ~+~ \de (F_n) ~-~ (-1)^n. \label{dn} \eeq
To solve this equation, we define
\beq \de' (F_{n+1}) ~=~ \de (F_{n+1}) ~+~ (-1)^n. \label{tdn} \eeq
Eq.~\eqref{dn} then implies the recursion relation
\beq \de' (F_{n+1}) ~=~ \de' (F_{n-1}) ~+~ \de' (F_n). \label{tdn2} \eeq
Using the fact that $\de' (F_2) = -1$ and $\de' (F_3) = 0$,
we find from Eq.~\eqref{tdn2} that
\beq \de' (F_{n+1}) ~=~ F_{n-2}, \eeq
which implies that
\beq \de (F_{n+1}) ~=~ - ~F_{n-2} ~-~ (-1)^n. \label{dn2} \eeq
This means that
\beq \frac{\de (N)}{N} ~=~ - ~\frac{1}{G^3} \label{asymp} \eeq
in the limit that $N = F_{n+1} \to \infty$.

We will now study what happens if $N$ is close to but not equal to $F_{n+1}$.
In particular, let us consider what happens if $N = F_{n+1} + j$, where 
$1 \le j \ll F_{n+1}$. We can then show that the corresponding unitary 
operator can be written as
\beq U (N) ~=~ V_{m_1} ~V_{m_2} ~\cdots~ V_{m_r} ~V_n, \label{form} \eeq
where the $r$ integers $m_j$ satisfy
\beq 1 ~\le~ m_1 ~<~ m_2 ~<~ \cdots ~<~ m_r ~\ll~ n. \label{const1} \eeq
Further, Eq.~\eqref{vn} implies that we can assume that no two successive 
integers, $m_j$ and $m_{j+1}$, are consecutive integers; otherwise, if 
$m_{j+1} = m_j + 1$, we could use Eq.~\eqref{vn} to re-write 
$V_{m_j} V_{m_j+1}$ as $V_{m_j+2}$. Hence we will assume that 
\beq m_j < m_{j+1} ~-~ 1 \label{const2} \eeq
for all $j=1,2,\cdots,r-1$. For the operator in Eq.~\eqref{form}, we see that
\beq N ~=~ F_{n+1} ~+~ \sum_{j=1}^r ~F_{m_j+1}. \label{totN} \eeq
Eq.~\eqref{const1} implies that the sum in Eq.~\eqref{totN} is much smaller 
than $F_{n+1}$; to leading order, therefore, we still have $N \simeq F_{n+1} 
\simeq G^{n+1}/\sqrt{5}$. 

We can now make repeated use of Eq.~\eqref{abd} to compute $\de (N)$. Since we 
are interested in calculating $\de (N)/N$ in the limit $N \to \infty$, we only 
need to keep terms of the form $\al (M) \beta (N) - \beta (M) \al (N)$, where 
$M$ can take the values $F_{m_1+1}, ~F_{m_2+1}, \cdots, F_{m_r+1}$ but $N = 
F_{n+1}$ is held fixed. All the other terms, where $M=F_{m_j+1}$ and 
$N=F_{m_k+1}$, give contributions which are much smaller than $F_{n+1}$.
We then obtain
\beq \de (N) ~=~ - ~F_{n-2} ~+~ \sum_{j=1}^r ~[\al (F_{m_j+1}) \beta (N) ~-~ 
\beta (F_{m_j+1}) \al (N)]. \label{dn3} \eeq
Next, we have
\bea \al (F_{m_j+1}) \beta (N) ~-~ \beta (F_{m_j+1}) \al (N) &=& 
F_{m_j} F_{n-1} ~-~ F_{m_j -1} F_n \non \\
&\simeq& \frac{G^{n+1}}{\sqrt{5}} ~\frac{(-1)^{m_j+1}}{G^{m_j+1}}, \label{dn4}
\eea
to leading order. Dividing this by $N \simeq G^{n+1}/\sqrt{5}$ gives
\beq \frac{\de (N)}{N} ~=~ - ~\frac{1}{G^3} ~+~ \sum_{j=1}^r ~
\frac{(-1)^{m_j+1}}{G^{m_j+1}}. \label{dn5} \eeq
This is one of our main equations.

The maximum and minimum values of the expression in Eq.~\eqref{dn5} can be 
found as follows. Noting that odd (even) values of $m_j$ make positive 
(negative) contributions, we see that the maximum value of $\de (N)/N$ arises 
for the case $m_1 =1, ~m_2 =3, ~m_3 = 5, ~\cdots$. 
(Note that this satisfies the constraint in Eq.~\eqref{const2}). We then get
\bea \left( \frac{\de (N)}{N} \right)_{max} &=& - ~\frac{1}{G^3} ~+~ 
\sum_{j=1}^\infty ~\frac{1}{G^{2j}} \non \\
&=& 1 ~-~ \frac{1}{G} ~\simeq~ 0.382. \label{dmax} \eea
(In taking the sum to go up to $j=\infty$ in the first line of 
Eq.~\eqref{dmax}, we have assumed that $n$ is very large so that $m_r \ll n$
remains consistent even for a large value of $m_r$). The minimum value of $\de
(N)/N$ arises for the case $m_1 =2, ~m_2 =4, ~m_3 = 6, ~\cdots$, which gives
\bea \left( \frac{\de (N)}{N} \right)_{min} &=& - ~\frac{1}{G^3} ~-~ 
\sum_{j=1}^\infty ~\frac{1}{G^{2j+1}} \non \\
&=& -~ \frac{1}{G} ~\simeq~ - ~0.618. \label{dmin} \eea
Note that the values given in Eqs.~\eqref{dmax} and \eqref{dmin} differ by 1.

\begin{figure}[ht]
\includegraphics[height=9.5cm]{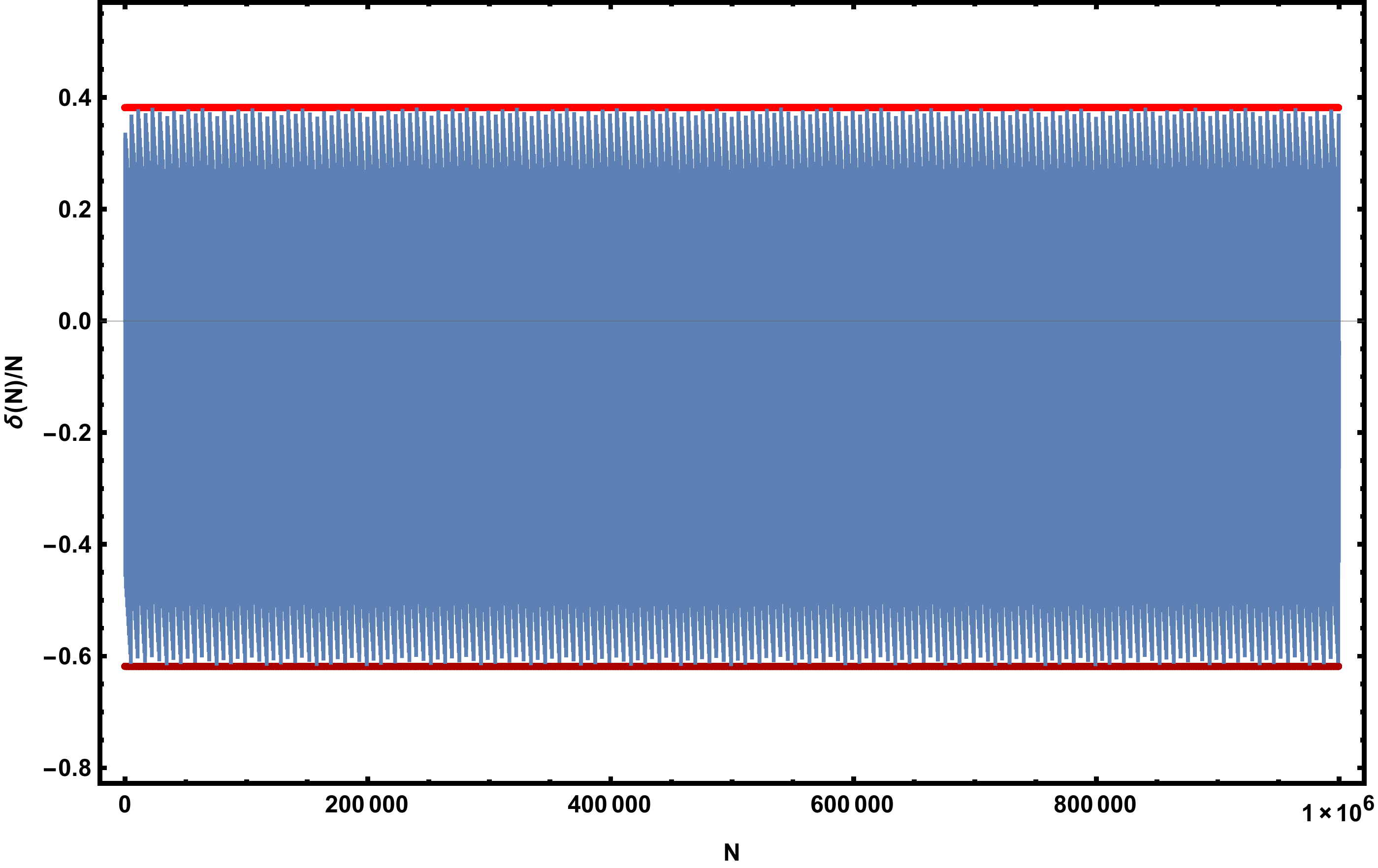}
\caption{Plot of $\de (N)/N$ versus the number of drives $N$. The horizontal 
red lines lie at the values $1 - 1/G \simeq 0.382$ and $-1/G \simeq -0.618$.} 
\label{delN} \end{figure}

Figure~\ref{delN} shows a plot of $\de (N)/N$ versus $N$ from 1 to $10^6$. We 
see that $\de (N)/N$ lies between the values $1-1/G$ and $-1/G$ which 
are shown by horizontal red lines. The fact that a large region between these 
two numbers seems to be filled up suggests that a very large number of values 
of $\de (N)/N$ can indeed be written in the form given in Eq.~\eqref{dn5}.

To summarize, we have shown that for a large number of drives, $N$, of the 
order of $F_{n+1} \simeq G^{n+1}/\sqrt{5}$, the unitary operator is given by
\bea U (N) &=& \exp[ \al (N) A ~+~ \beta (N) B ~+~ \de (N) ~[A, B]/2], \non \\
{\rm where} ~~~\al (N) &\simeq& \frac{G^n}{\sqrt 5},~~~ 
\beta (N) ~\simeq~ \frac{G^{n-1}}{\sqrt 5},~~~{\rm and}~~~
\de (N) ~\simeq~ \ga \frac{G^{n+1}}{\sqrt 5}, \label{un3} \eea
where $\ga$ in the last equation fluctuates between the values $-1/G$ and
$1 - 1/G$. We now consider the case where $A$ and $B$ are equal to 
$i$ times some linear combinations of the three Pauli matrices. We 
can then write Eq.~\eqref{un3} in the form
\beq U(N) ~=~ e^{i \ta_N {\hat e}_N \cdot {\vec \si}}, \label{un4} \eeq
where the angle $\ta_N$ and the unit vector $\hat e_N$ vary with $N$;
we can think of $U(N)$ as performing a spin rotation by an angle $2\ta_N$
around the axis ${\hat e}_N$. We now 
assume that $[A,B]$ is much smaller than $A$ and $B$; this is true if 
the stroboscopic time period $T$ is small since $A$ and $B$ are proportional 
to $T$ implying that $[A,B]$ is proportional to $T^2$. We then see from 
Eq.~\eqref{un3} that ${\hat e}_N \cdot {\vec \si}$ 
is almost fixed and is proportional to $i(G A + B)$ (multiplied by a 
normalization factor to ensure that its eigenvalues are equal to $\pm 1$),
while $\ta_N$ varies with $n$ as $G^n/{\sqrt 5}$. (There will be small 
fluctuations in ${\hat e}_N$ due to the term $\de (N) ~[A, B]/2$ as quantified
in Eqs.~\eqref{dmax} and \eqref{dmin}). Since the ratio of $G^n/{\sqrt 5}$ to 
$2\pi$ is an irrational number, $\ta_N$ modulo $2 \pi$ 
will cover the range $[0,2\pi]$ uniformly as $N$ varies. Thus the sequence
of unitary matrices $U(N)$ is described by a unit vector ${\hat e}_N$ which is 
almost fixed (except for small fluctuations) and an angle $\ta_N$ which, modulo
$2\pi$, takes all possible values in the range $[0,2\pi]$. 

\begin{figure}[ht]
\includegraphics[height=9.5cm]{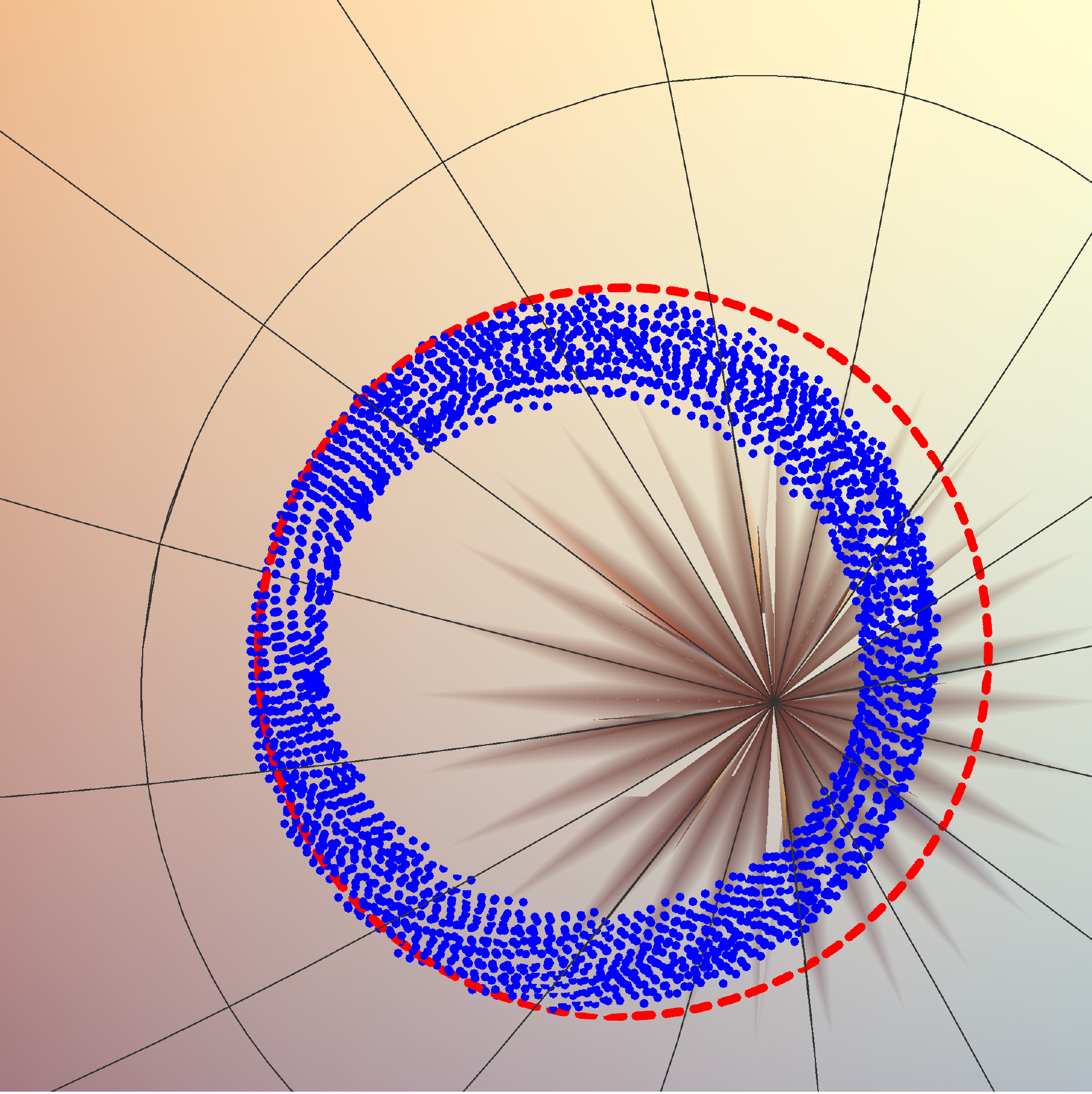}
\caption{This figure shows the zoomed out version of Fig. 1 (b) of the main 
text; we see more clearly the trajectory of the points on the Bloch sphere.} 
\label{fig_bss2} \end{figure}

We now consider what happens when a sequence of such matrices $U(N)$ acts on 
an initial state which is given by
\beq \ket{\psi (0)} ~=~ \begin{bmatrix}
\cos\left(\theta (0)/2\right) \\
\sin\left(\theta (0)/2\right) e^{i\phi (0)} \\
\end{bmatrix}. \eeq
This state can be mapped to a point on the Bloch sphere whose Cartesian 
coordinates are given by $z_0 = \cos (\theta (0)/2), ~x_0 = \sin (\theta (0)/2)
\cos (\phi (0))$ and $y_0 = \sin (\theta (0)/2) \sin (\phi (0))$. These 
coordinates define an initial unit vector ${\hat r}_0 = (x(0),y(0),z(0))$. 
After $N$ drives, we obtain
\beq \ket{\psi (N)}= U (N) \ket{\psi (0)} = \begin{bmatrix}
\cos\left(\theta (N)/2\right) \\
\sin\left(\theta (N)/2\right) e^{i\phi (N)} \\
\end{bmatrix}, \eeq
which corresponds to a point on the Bloch sphere with coordinates
$z_N = \cos (\theta (N)/2), ~x_N = \sin (\theta (N)/2) \cos (\phi (N))$, 
and $y_N ~=~ \sin (\theta (N)/2) \sin (\phi (N))$ which define
a unit vector ${\hat r}_N$. Given the form of $U(N)$ 
in Eq.~\eqref{un4}, where ${\hat e}_N$ is almost fixed while $\ta_N$ covers 
the range $[0,2\pi]$ as $N$ varies, we now see that the points given by 
${\hat r}_N$ move on a circle which rotates around the direction ${\hat e}_N$,
and the angle between ${\hat e}_N$ and ${\hat r}_N$ is almost fixed and is 
given by $\cos^{-1} ({\hat e}_N \cdot {\hat r}_0)$. This explains Fig. 3 in the
main text (see also Fig.~\ref{fig_bss2}) where we see that all the points 
almost lie on a circle. More precisely, the points lie between two nearby 
circles which are defined by the maximum and minimum values of $\de (N)/N$; the
separation between the two circles is of the order of $[A,B] \propto T^2$.

\section{Does the steady state remain the same up to arbitrarily large $N$ ?}

In this section, we will try to understand if the steady state (in which the
trajectory on the Bloch sphere is restricted to a thin ring which is centered 
around a particular axis) remains the same up to arbitrarily large values of 
$N$. We will argue that if the time period $T$ is small, the direction of the
axis changes by an amount $\ep$ after an astronomically large number of 
drives given by $N_0 \sim G^{\ep /T^2}$.
 
We will do our analysis for values of $N= F_{n+1}$ which are Fibonacci numbers.
The time evolution operators $V_n = U (F_{n+1})$ then satisfy the
recursion relation in Eq.~\eqref{vn} for $n \ge 2$. Let us parameterize
\beq V_n ~=~ e^{i \ta_n {\hat e}_n \cdot {\vec \si}} ~=~ \cos \ta_n ~I_2 ~+~ i 
\sin \ta_n ~{\hat e}_n \cdot {\vec \si}, \label{vn2} \eeq
where $I_2$ is the $2 \times 2$ identity matrix, and
$\ta_n$ can be taken to lie in the range $[0,2\pi]$.
It was shown by Sutherland~\cite{sutherland86} that the recursion relation
in Eq.~\eqref{vn} has an invariant; namely, the quantity
\beq S ~=~ \cos^2 \ta_n ~+~ \cos^2 \ta_{n+1} ~+~ \cos^2 \ta_{n+2} ~-~ 2 
\cos \ta_n \cos \ta_{n+1} \cos \ta_{n+2} ~-~ 1 \label{suth1} \eeq
is {\it independent} of $n$. Further, if $\Ga_{n,n+1} = \cos^{-1} ({\hat e}_n 
\cdot {\hat e}_{n+1})$ denotes the angle between ${\hat e}_n$ and 
${\hat e}_{n+1}$, the Sutherland invariant is equal to
\beq S ~=~ - \sin^2 \ta_n ~\sin^2 \ta_{n+1} ~\sin^2 \Ga_{n,n+1}. \label{suth2} 
\eeq

We now note that in the expressions
\beq V_0 ~=~ e^B ~~~{\rm and}~~~ V_1 = e^{A}, \eeq
$A = -i H^A T$ and $B = - i H^B T$ are both of order $T$. We can write
$H^A = {\vec a} \cdot {\vec \si}$ and $H^B = {\vec b} \cdot {\vec \si}$, where
the vectors $\vec a$ and $\vec b$ and therefore the angle between them are all 
of order 1. Looking at the Sutherland invariant $S$ in Eq.~\eqref{suth2} for 
$n=1$, we then see that $\sin \ta_1$ and $\sin \ta_2$ are of order $T$ while
$\sin \Ga_{n,n+1}$ is of order 1. Hence $S$ is of order $T^4$.

We now consider what happens if we use the recursion relation in Eq.~\eqref{vn}
many times. We found above that $\ta_n$ varies as $G^n$ modulo $2\pi$ for 
large $n$ and is therefore generally of order 1; hence $\sin \ta_n$ and $\sin 
\ta_{n+1}$ are also of order 1. Then the invariant in Eq.~\eqref{suth2} 
implies that 
$\sin^2 \Ga_{n,n+1}$ is of order $T^4$; hence the angle $\Ga_{n,n+1}$ 
between ${\hat e}_n$ and ${\hat e}_{n+1}$ must be of order $T^2$. Since $T$ 
has been assumed to be small, this explains why ${\hat e}_n$ changes very 
slowly with $n$. Since ${\hat e}_n$ changes by only order $T^2$ when $n$ 
increases by 1, $n$ must increase by $\ep c/T^2$ for ${\hat e}_n$ to change by 
an angle $\ep$ (here $c$ is a number of order 1 which depends on the values of
$\vec a$ and $\vec b$ appearing in $H^A$ and $H^B$). Since $N= F_{n+1} \simeq 
G^{n+1}/ \sqrt{5}$ for large $n$, we see that $N$ must reach a value of the 
order of $N_0 \sim G^{\ep c/T^2}$ before ${\hat e}_n$ (which determines 
the axis around which the ring in the Bloch sphere is centered) will change by 
order $\ep$. For small values of $T$, this is an astronomically large number.
In our numerical calculations, we have taken $\om = 500$ so that $T=2\pi/500$.
Hence $N_0 \sim 1.618^{\ep c (500/2\pi)^2}$ which is about $10^{13}$ for $\ep = 
0.01$ and $c=1$. We thus see that the axis of rotation on the Bloch sphere 
can change appreciably and therefore give a different state, but this can 
only occur after an extremely large number of drives which would not be
discernible within experimental time scales.

\end{document}